# Actionable Principles for Artificial Intelligence Policy
## Three pathways

Charlotte STIX[1]

Eindhoven University of Technology, The Netherlands.

**Abstract:** In the development of governmental policy for artificial intelligence (AI) that is informed by ethics, one avenue currently pursued is that of drawing on "AI Ethics Principles". However, these AI Ethics Principles often fail to be actioned in governmental policy. This paper proposes a novel framework for the development of 'Actionable Principles for AI'. The approach acknowledges the relevance of AI Ethics Principles and homes in on methodological elements to increase their practical implementability in policy processes. As a case study, elements are extracted from the development process of the *Ethics Guidelines for Trustworthy AI* of the European Commission's "High Level Expert Group on AI". Subsequently, these elements are expanded on and evaluated in light of their ability to contribute to a prototype framework for the development of 'Actionable Principles for AI'. The paper proposes the following three propositions for the formation of such a prototype framework: (1) preliminary landscape assessments; (2) multi-stakeholder participation and cross-sectoral feedback; and, (3) mechanisms to support implementation and operationalizability.

---

[1] PhD Candidate, Philosophy and Ethics Group, Department of Industrial Engineering and Innovation Sciences, Eindhoven University of Technology, The Netherlands. Correspondence should be addressed to c.stix@tue.nl. ORCID iD: 0000-0001-5562-9234. During the writing of this paper the author acted as Coordinator of the European Commission's High Level Expert Group on Artificial Intelligence. The author neither discloses confidential information nor makes use of information obtained during that work in this paper.

## Introduction

Recent years have seen a veritable surge of Ethics Principles[2] for artificial intelligence[3] (AI) (Fjeld et al., 2020; Hagendorff, 2020; Ryan & Stahl, 2020; Jobin et al., 2019; Zeng et al., 2018; Morley et al., 2019). This in turn led to critical debates over the usefulness and impact of such instruments, with a particular focus on what is often held as a lack of implementation of such AI Ethics Principles into actual policy-making. In response, this paper will propose a preliminary framework to support and improve the implementation of AI Ethics Principles in governmental policy at this critical time.

The appeal of AI Ethics Principles lies in their promise to condense complex ethical considerations or requirements into formats accessible to a significant portion of society, including both the developers and users of AI technology. To live up to this, however, these principles face two high-level challenges: (a) they must achieve a succinct condensation of broad and deep ethical theories into an accessible number of principles, and (b) they must strike a balance between pursuing an ideal hypothetical outcome, and working to secure workable pragmatic outcomes. In doing so, it is important to recognize that while 'workable pragmatic outcomes' may rightly be perceived as suboptimal from a strict ethical perspective, they will often form the critical basis to moving AI Ethics Principles forwards into practical policy.

This paper focuses on helping AI Ethics Principles strike such a balance, by complementing previous work (which focuses largely on identifying the ideals to be pursued) with a perspective that enables the achievement of 'workable pragmatic outcomes' in AI policy. To do so, it proposes one avenue to increase AI Ethics Principles' operationalizability into policy, ensuring these are actionable for governmental actors. It therefore limits itself to exploring one particular bottleneck that AI Ethics Principles face, and does not claim to resolve other, equally pressing shortcomings. In short, the goal is not to explore what precisely should be *in* AI Ethics Principles per se, but examine one angle as to *how* they should be developed to be actionable in a specific domain.

---

[2] For the purpose of this paper, "AI Ethics Principles" encompass all documents outlining policy for the development
and deployment of AI, based on ethical considerations.

[3] For the purpose of this paper, the author follows the definition of AI of the European Commission (European Commission, 2018, p.1) stating that AI systems "display intelligent behaviour by analysing their environment and taking actions – with some degree of autonomy – to achieve specific goals".

This paper puts forward an initial framework for the development of what it calls *Actionable Principles for AI Policy*. To do so, it will proceed as follows: in section "Case Study: The Ethics Guidelines for Trustworthy Artificial Intelligence" it will present three key procedural elements of the 'Ethics Guidelines for Trustworthy Artificial Intelligence' (henceforth: 'Ethics Guidelines', AI HLEG 2019b) presented in 2019 by the European Commission's independent High Level Expert Group on Artificial Intelligence (AI HLEG). Reviewing these procedural instruments, this paper will subsequently expand and build thereon. On this basis, section "A Preliminary Framework for Actionable Principles" will culminate in a proposal for an initial framework for Actionable Principles for AI.

## Actionable Principles

In many areas, including AI, it has proven challenging to bridge ethics and governmental policy-making (Müller 2020, 1.3). To be clear, many AI Ethics Principles, such as those developed by industry actors or researchers for self-governance purposes, are not aimed at directly informing governmental policy-making, and therefore the challenge of bridging this gulf may not apply. Nonetheless, a significant subset of AI Ethics Principles are addressed to governmental actors, from the 2019 OECD Principles on AI (OECD 2019) to the US Defence Innovation Board's AI Principles adopted by the Department of Defence (DIB 2019). Without focussing on any single effort in particular, the aggregate success of many AI Ethics Principles remains limited (Rességuier and Rodriques 2020). Clear shifts in governmental policy which can be directly traced back to preceding and corresponding sets of AI Ethics Principles, remain few and far between. This could mean, for example, concrete textual references reflecting a specific section of the AI Ethics Principle, or the establishment of (both enabling or preventative) policy actions building on relevant recommendations. A charitable interpretation could be that as governmental policy-making takes time, and given that the vast majority of AI Ethics Principles were published within the last two years, it may simply be premature to gauge (or dismiss) their impact. However, another interpretation could be that the current versions of AI Ethics Principles have fallen short of their promise, and reached their limitation for impact in governmental policy-making (henceforth: policy).

It is worth noting that successful actionability in policy goes well beyond AI Ethics Principles acting as a reference point. Actionable Principles could shape policy by influencing funding decisions, taxation, public education measures or social security programs. Concretely, this could mean increased funding into societally relevant areas, education programs to raise public awareness and increase vigilance, or to rethink retirement structures with regard to increased automation. To be sure, actionability in policy does not preclude impact in other adjacent domains, such as influencing codes of conduct for practitioners, clarifying what demands workers and unions should pose, or shaping consumer behaviour. Moreover, during political shifts or in response to a crisis, Actionable Principles may often prove to be the only (even if suboptimal) available governance tool to quickly inform precautionary and remedial (legal and) policy measures.

There exist concrete examples demonstrating that some select AI Ethics Principles already do possess a degree of actionability. For instance, the Ethics Guidelines (AI HLEG 2019b) have had a significant policy impact within the European Union (EU). They influenced both the political Agenda of Commission President Von der Leyen (2019) and informed the initial legislative framework proposal for AI. The latter used the Ethics Guidelines' seven key requirements for 'trustworthy AI'[4] to form the basis for the legal obligations which any 'high-risk' AI system would need to fulfill in order for it to be deployed within the EU (European Commission 2020). Arguably, these Ethics Guidelines were among the chief documents available to inform and guide the content of this legislative proposal. The aforementioned highlights the potential

---

[4] Defined by the AI HLEG as being (i) ethical, (ii) lawful, and (iii) robust from a socio-technical perspective.

importance—and promise—of how future Actionable Principles could be set to significantly shape the development, deployment and use of AI by virtue of their influence on policy.

This paper will inevitably touch on a variety of concerns that afflict AI Ethics Principles beyond their lack of actionability. These are often intertwined with (if not the result of) procedural shortcomings that affect actionability. Some of these are, for example, a *lack of clarity* which can contribute to divergent interpretations (Whittlestone et al. 2019) or a *lack of balanced participation* which can contribute to 'ethics washing' (Floridi et al. 2018). The latter practice is commonly alleged with regards to industry-driven AI Ethics Principles, and can result in superficial proposals that mask themselves as ethical, but may, in fact, be commercially or politically motivated (Floridi et al. 2018, p. 187).

Finally, it should be noted that underlying all this, there are a multitude of parallel discussions about the best governance approaches towards AI—and whether novel AI Ethics Principles are even the best tool in the first place. For instance, one prominent proposal is to use the international human rights framework as the basis for setting the ground for ethical AI systems. Indeed, this tool could be a valuable angle given the existing legitimacy and consensus this approach can draw from. Nevertheless, this paper takes a different angle and focuses on improving the ability of AI Ethics Principles to shape a given governance framework. It does not make a judgement on the relative value of various approaches in AI governance, but instead focuses on the improvement of one specific approach. Current AI Ethics Principles have been critiqued for the fact that, while they "may guide the entities that commit to them, […] they do not establish a broad governance framework." (Donahoe and Metzger 2019, p. 118). To that end, this paper proposes a concept of 'Actionable Principles'. Inevitably, some suggestions for Actionable Principles will draw on what has been successful in the past, and therefore stands to be promising in the future. These procedural elements, such as proposals for multi-stakeholder and cross-sectoral dialogue (Donahoe and Metzger 2019; Yeung et al. 2019), are likely to overlap with, for example, ongoing suggestions for the establishment of an international human rights based framework for the purpose of AI governance. Building a framework for policy-effective Actionable Principles for AI is therefore not meant to be in competition with other approaches, but rather to complement them.

## 1. Case study: The Ethics Guidelines for Trustworthy Artificial Intelligence

This section focuses on the development process of the AI HLEG's Ethics Guidelines (2018–2019), in order to highlight three promising procedural elements, which will be subsequently developed in section "A Preliminary Framework for Actionable Principles". The Ethics Guidelines are selected because they arguably advanced the state-of-the-art of AI Ethics Principles by virtue of directly informing policy-making within the EU (European Commission 2020), and because they are grounded in the protection of fundamental rights (AI HLEG 2019a, b; AI HLEG 2020). The Ethics Guidelines therefore constitute a promising case to draw transferable or generalizable lessons from for Actionable Principles.

Of course, it has to be noted that neither the Ethics Guidelines nor their development process are void of criticism. Various critiques have been raised to it, from their alleged development under outsized industry influence (Hidvegi and Leufer 2019; Article 19 2019) and the obfuscation of 'red lines' (Metzinger 2019; Klöver and Fanta 2019), to a lack of matching governance structures to achieve real impact (BEUC 2019; Veale 2020). As such, it is important to mark that although the framework for Actionable Principles proposed in this paper is informed by certain procedural elements of the Ethics Guidelines, it makes no assumption about the relative value of existing criticism of the Ethics Guidelines as a whole.

### 1.1 Diversity

The AI HLEG was a large group of experts from multiple sectors, ranging from ethicists, lawyers, to machine learning researchers, trade unionists, and various other stakeholders.[5] This diversity allowed the AI HLEG to provide informed recommendations sensitive to a variety of concerns. Moreover, the AI HLEG benefited from continuous engagement (including on earlier drafts of the Ethics Guidelines) with the European AI Alliance, a multi-stakeholder platform with over 4000 entities and various subject experts across Europe.[6] Notwithstanding that, there is room for improvement in the establishment of groups developing Actionable Principles, especially in light of concerns over dominant representation of industry friendly voices (Hidvegi, Leufer, 2019) and a lack of sufficient representation of AI Ethicists (Metzinger, 2019) within the AI HLEG. While the European Commission has a strict selection process with different types of membership criteria during open calls,[7] it could, in the future, improve a lack of representation especially from civil society voices such as the European Network Against Racism, by explicitly hand-selecting them outside of public calls. Such a procedure could be within the remit of considering it an "overriding priority" that certain groups are adequately represented in given expert groups through membership.

---

[5] https://ec.europa.eu/digital-single-market/en/high-level-expert-group-artificial-intelligence

## 1.2 Working methods

The AI HLEG solicited public feedback not once, but twice, both during the development process, but also after publishing the final document. The first public consultation concerned the draft Ethics Guidelines, made available for public feedback roughly half a year into the writing process. Such an intermittent solicitation, though beneficial to inform the development process, is not the norm. The vast majority of groups tasked with the development of AI Ethics Principles solicit feedback ex ante (if at all). Generally, little to no leeway is given for amendments during the writing process, although some other notable exceptions can be found in e.g. the Australian AI Ethics Framework (DISER 2019) or the New Zealand Draft Algorithm Charter (New Zealand Government 2020). With the AI HLEG, this dynamic procedure allowed for amendments of initial propositions and engaged stakeholders to co-create an ethical framework, taking their concerns, unique expertise and suggestions into consideration.

Subsequently, the AI HLEG presented revised Ethics Guidelines in April 2019. Shortly thereafter, the AI HLEG continued to work in an agile manner, stress-testing the suitability of their recommendations in the real world. More precisely, in June 2019, two months post-publication of their final Ethics Guidelines, the European Commission opened a 'piloting phase' on behalf of the AI HLEG. This phase concerned itself with the third section of the Ethics

---

[6] https://ec.europa.eu/digital-single-market/en/european-ai-alliance
[7] https://ec.europa.eu/transparency/regexpert/index.cfm?do=faq.faq&aide=2

Guidelines, which contained an assessment list meant to support the actionability of the Ethics Guidelines' key requirements, i.e. the main recommendations. As such, this piloting phase was meant to trial the usefulness, comprehensiveness and suitability of this list. The goal was to receive feedback that would allow the AI HLEG to improve the assessment list encouraging better actionability (AI HLEG 2020). Feedback was solicited through a three-pronged approach: (i) via 50 in-depth day long interviews with selected companies; (ii) two quantitative surveys for technical and non-technical stakeholders; and, (iii) a dedicated space on the AI Alliance where feedback could be submitted. Together, this allowed for a breadth of multidimensional input.

## 1.3 Toolboxes

The 'toolboxes' of mechanisms accompanying the Ethics Guidelines enabled operationalisability of the key requirements. These toolboxes contained both technical and non-technical methods and recommendations which could be used independently, simultaneously, or, consecutively. The technical toolbox proposed to make use of e.g. architectures for trustworthy AI, testing and validation methods, explainable AI (XAI) research, as well as Quality of Service indicators. The non-technical toolbox ranged from public awareness and diversity measures, to efforts that can be undertaken by governments such as standardisation, certification, and regulation. The fact that the AI HLEG included non-technical methods demonstrates that the responsibility to co-create, maintain and deploy trustworthy AI extends well beyond the technical arena. Similarly, the range of methods proposed reflected the range of stakeholders that are necessary to develop 'trustworthy AI': from governments with the mandate to create regulation, to researchers with their ability to shape the implications and features of their AI development process, and from industry actors who are in a position to create more diverse hiring processes (Crawford 2016), to civil society actors with their power to demand and engage in multi-stakeholder dialogues.

In summary, the AI HLEG process highlighted the following procedural mechanisms: (i) involvement of diverse voices, both between the experts and through open public feedback; (ii) agile development process through interim and ex post feedback processes; and, (iii) mechanisms and methods to support actionability. These aspects appear particularly promising components for grounding and supporting an initial approach to developing Actionable Principles.

## 2. A preliminary framework for Actionable Principles

The above brief review of the development process of the Ethics Guidelines indicates three promising elements: support of diversity, allowance for agile development and support for implementation. This section builds on these elements, in order to generalize them and expands their scope. It proposes a preliminary framework towards Actionable Principles composed of the following: (1) preliminary landscape assessments; (2) multi-stakeholder participation and cross-sectoral feedback; and (3) mechanisms to support implementation and operationalizability.

These steps cover crucial turning points at each stage of the development process towards Actionable Principles, from inception to development, to their post-publication stage. A (1) *preliminary landscape assessment* addresses the contextual environment within which Actionable Principles arise. Once that has been established, (2) *multi-stakeholder participation and cross-sectoral feedback* addresses the composition and working methods of those parties which draft any sets of Actionable Principles. Finally, (3) *support of implementation and operationalizability* addresses the direct move towards Actionable Principles' implementation into governmental policy-making post publication. In discussing each of the three elements of this prototype framework in turn, it can be seen how they also relates to - or provides insights on - some of the common critiques of existing AI Ethics Principles.

### 2.1 Development of preliminary landscape assessments

Actionable Principles benefit from what this paper calls a 'landscape assessment'. This could inform their development process, and serve to place Actionable Principles within the particular environment (geopolitical, societal, legal etc.) in which they are implemented, as well as to identify blindspots or practical difficulties before they arise. In a similar manner to government departments comparing various policy options in order to decide on the best one to implement, or to conduct impact assessments prior to introducing new regulation, landscape assessments can support foresight and analysis in the drafting of Actionable Principles, making them more impactful and applicable.

Landscape assessments therefore create a bridge between what should ideally be done, and what can be done (and how), supporting a step change into the right direction. Promising areas for a landscape assessment include the technical state-of-the-art, identifying which capabilities and applications exist, what societal, economic and political factors may affect their potential proliferation or market penetration to various actors, and the resulting timelines of sociotechnical change; such assessments also include the societal environment, to determine what are the public's and policymaker's overall understanding, range of concerns, and ability to engage with issues. Finally, it could serve to review the legislative status quo, to understand the scope of issues already covered (Gaviria 2020).

Overall, landscape assessments are likely to concretise and strengthen an ethical principle such as 'an AI should not unduly influence human agency' which otherwise might be too broad (that is, not specific enough about its requirements), or too overarching in the scope of the AI applications it seeks to apply to in order for it to be actionable and functional in policy. For example, a landscape assessment could address the technical state-of-the-art for synthetic media, identifying a lack of technical tools to adequately capture all synthetic audio and visual products, and point towards civil society being insufficiently educated about this technological capability. A landscape assessment might also identify that the Californian B.O.T. Act (Bolstering Online Transparency Act 2018) requires bots on the internet to self-identify in order to not mislead humans, however, highlight that this is not generally applicable to all 'output' derived from an AI agent.

Similarly, an assessment of the state-of-the-art of energy-efficient learning in AI, and tradeoffs between beneficial AI applications now versus the impact of climate change going forwards, would have value. For example, this could ensure that ethical considerations for future generations or tensions between developing technological solutions now at a potential longer-term cost can be clearly flagged and evaluated in advance. In short, a landscape assessment could support the identification of practical issues that map onto more theoretical aspects, facilitating more actionable policy. Actionable Principles informed by landscape assessments can serve to form the backbone for timely and evidence-based policy making. Moreover, they address a common criticism that AI Ethics Principles lack access to adequate information necessary to make impactful recommendations.

A lack of adequate and pertinent information, or the ability, resources and authority to gather such, underlying the development process of any set of AI Ethics Principles, can jeopardize the resulting impact significantly. In particular, it could hinder the ability to foresee practical second-order issues that may result out of the recommendations provided, or to make truly impactful and actionable recommendations.

Looking back over the past years, various cases illustrate how a lack of advanced landscape assessments can constrain the downstream impact of AI Ethics Principles. In 2017, New York City enacted bill "Int. 1696", establishing an Automated Decision Systems Task Force (ADS). The goal of this task force was to develop recommendations reviewing New York City's use of automated decision-making systems, including individual instances of "harm". However, their final report was subject to critique, which amongst other points homed in on its weak recommendations, which were partially traced back to the lack of resources the ADS had initially been allocated to conduct an appropriate landscape assessment. On this basis, AI Now Institute's Shadow Report on the ADS's work (Richardson 2019) emphasized that any task force or group meant to "review, assess and make recommendations" should be empowered to receive all necessary information to the fulfillment of their task in order to be able to appropriately evaluate the landscape and provide informed outputs. This includes access to existing laws, policies and guidelines. Going forwards, stakeholder groups drafting Actionable Principles need to be able to ensure that the impact of their work is not minimised by a lack of, or, ability to conduct a landscape assessment.

A landscape assessment could address practical hurdles of actionability that result out of a lack of access to sufficiently comprehensive and relevant information. It therefore constitutes the first pillar in formulating Actionable Principles for AI. However, while necessary, it is not sufficient by itself to ensure actionability. For this, we must turn to additional mechanisms.

## 2.2 Multi-stakeholder participation and cross-sectoral feedback

In order to achieve an outcome that works for all, the stakeholder group developing Actionable Principles must be diverse and representative. This allows for a broad (and relevant) range of concerns and recommendations to be taken into account and accurately reflected. After all, one goal of Actionable Principles is to encourage good policy-making that works for all of society. Despite being a relatively diverse multi stakeholder group, the AI HLEG's composition could have been improved, e.g. when it came to the balance between civil society and industry representatives (Hidvegi and Leufer 2019). Indeed, during the composition of multi-stakeholder groups, special care should be taken that those who are most likely to be adversely affected due to the development, deployment and application of certain AI systems are adequately represented, have their opinions heard, and can have an outsized say on the issue at hand.
Intra-group diversity is necessary, it is not sufficient. Instead of focussing on the ideal building blocks for intra-group diversity, this paper will explore an expansion to this goal. For Actionable Principles this could take the form of significant agile engagement with multiple cross-sectoral stakeholders outside of the group, consulting them on their expertise, insights and unique point of view. Elsewhere, such an approach has also been proposed for 'human rights-centred design, deliberation and oversight' of AI (Yeung et al. 2019). It ensures that a much larger diversity of voices is captured that otherwise would be missing in the process, regardless of the composition or size of the group. This is an important consideration for the development of Actionable Principles because no group, no matter how diverse, can possibly reflect the knowledge and expertise of hundreds of diverse stakeholders. Moreover, shortcomings such as group think are more likely when groups are unable to actively engage with outside stakeholders.

The predominant method of soliciting external expertise is to conduct a public consultation. Broadly speaking, there appear to be three different approaches: (i) consultations prior to the drafting stage; (ii) consultations during the drafting stage; and, (iii) consultations post publication stage.

Most commonly, (i) consultations are conducted prior to the drafting stage. These consultations often take the form of either presentations to the stakeholder group tasked with the drafting, or the submission of written position papers. These submissions are often based on a set of questions meant to inform and guide their content. For example, the report of the Select Committee on AI from the House of Lords (House of Lords 2018) was written based on feedback received in this manner.

While not replacing a comprehensive landscape assessment, an ex ante consultation can act to inform of the concerns, suggestions and existing difficulties that multiple stakeholders experience or foresee in the nearer future on a range of topics. Indeed, in the context of broader technology policymaking, the paradigm proposed by the Responsible Research and Innovation Framework has long emphasized such consultations (Owen et al. 2013; Stilgoe et al. 2013), suggesting they should play a crucial role in structural frameworks for the development of Actionable Principles. At the same time, ex ante consultations, particularly without a preliminary landscape assessment to guide them, suffer from shortcomings that could negate their usefulness. In particular, they often shift the burden of knowing what could constitute 'useful' information on the person or entity submitting information. They equally force the stakeholder group soliciting that information to already have a clear sense of the type of information they need, by virtue of asking the 'correct' questions or call for input. This can lead to two issues that constitute reverse sides of the same coin: those submitting may submit overly generic feedback, and the stakeholder group may ask overly broad questions. Both ultimately lead to types of feedback that may end up lacking (or swamping) the information that would have been relevant in hindsight.

A consultation prior to the drafting stage would therefore benefit from building on a preliminary landscape assessment. This would increase the clarity as to what type of additional information is sought from stakeholders and ensure that time on submissions is well spent, especially bearing in mind that some groups may not be able to spend significant amounts of time on submissions. This type of consultation would also benefit from being supplemented by the second and third type of consultation—an (ii) intermittent consultation during the drafting stage and a (iii) consultation post-publication stage, ensuring a flexible development process and timely feedback loop. Unfortunately, both (ii) and (iii) are rarely if ever made use of. The following paragraphs will outline why Actionable Principles would, however, benefit from these tools.

An (ii) intermittent consultation during the drafting stage can serve to stress test an initial proposal and steer the work towards actionability, in accordance with valid critiques and additional information derived from the consultation. Such a dynamic feedback loop can serve to elevate work from expert led to multi-stakeholder led with crowdsourced input. In addition to such an intermittent consultation on the Ethics Guidelines, stakeholders had the opportunity to engage directly with the AI HLEG via the European AI Alliance. Elements such as these could be useful to increase the comprehensiveness of Actionable Principles.

The third avenue of consultation, (iii) post publication stage, likewise remains missing from most published AI Ethics Principles. Such a consultation could assist in understanding reasons for a potential lack of implementation, or pinpoint omissions, outdated assumptions, or areas that require revision since the moment of publication. This type of consultation could support either a revision of the outcome or serve as a guiding structure for future attempts. For the purpose of achieving Actionable Principles, the former would be preferable as it has the potential to refine and adapt the document over time, strengthening it and encouraging the interplay between ethics and agile policy making.

All three avenues for consultation support the inclusion of different types of expertises, cross-sectoral dialogues and understandings, while counteracting a potential 'ivory tower mentality' which a stakeholder group closed to engagement with those external to the group may easily fall prey to.

One angle of concern, or criticism, is whether existing AI Ethics Principles accurately capture public and expert led-consensus. It should be noted that on many topics, consensus may be impossible to reach despite broad stakeholder inclusion. Green (2018) argues that there are certain topics which may wrongly be perceived as a constant, when they are actually a dynamic social construct. Take, for example, 'safety'. In an analysis by Fjeld et al. (2020) of AI Ethics Principle-type documents, over 81% contained some recommendations in favour of 'safety' and 'security'. Despite this apparent high-level commonality and agreement, they all suffer from a major deficit in light of Green's (2018) argument. Specifically, he argues that perceptions of what constitutes safety thresholds differ amongst people and societies. This would entail that e.g. AI would need to pass the median between all individual safety thresholds distributed across society, in order to be perceived as safe even if the concept is understood by all to denote the same thing. Whilst some ethical concerns may intuitively be perceived to have 'fuzzy boundaries', few would say the same about a concept such as 'safety'. Moreover, Jobin et al. (2019) identify significant variations in the different interpretations of commonly shared principles such as transparency. The likelihood of 'fuzzy boundaries' embedded in recommendations of Actionable Principles, be that on safety, environmental wellbeing, transparency or others, demonstrates the overall importance of introducing a multi-stakeholder approach reflecting as many considerations and concerns as possible. This can support Actionable Principles to avoid baking in 'suitcase terms' that lend themselves to diverse interpretation and future contestation.

Building on the aforementioned point, various groups, depending on their background, can end up using the same terminology, yet denoting different content spaces resulting in varying interpretations for their resolution. This creates the appearance of agreement at the drafting stage while causing difficulties at the implementation stage. For example, Xiang and Raji (2019) analysed various ways in which the technical research community makes use of legal terminology around fairness, but often lacks the necessary legal understanding to align their interpretation with what fairness is assumed to mean from a legal point of view, and vice versa. This misalignment between terminology and interpretation can quickly become a practical policy problem. Moreover, it can yield methodologically different strategies to address what is perceived as the same recommendation at hand (Whittlestone et al. 2019). The same recommendation may be seen as a political, ethical or technical problem, depending on the stakeholders involved trying to resolve it. Finally, this may introduce questions whether or not data-driven or algorithmic approaches should be outright banned in certain situations where individuals stand to be harmed by a subpar solution, or where the technical capability is not set to meet the requirements necessary from a societal and legal point of view (Wachter et al. 2020).[8]

In conclusion, in order to support Actionable Principles ongoing multi-stakeholder debate and broad continuous information exchange via e.g. multiple public consultations is key.

---

[8] This section does not aim to resolve whether or not fairness can be automated but how stakeholder dialogue could serve to identify tensions between different stakeholders' opinions and approaches.

## 2.3 Mechanisms to support implementation and operationalizability

Most AI Ethics Principles are drafted without concrete plans for their implementation, impact or target audience (Schiff et al. 2019). Indeed, they are often abstract and make vague suggestions (Mittelstadt 2019). This can hinder their actionability and implementation in policy-making. Guidance in the form of a toolbox, or method to operationalise the recommendations can be a crucial step to move from AI Ethics Principles towards Actionable Principles (Morley et al. 2019). The benefit of providing existing and desired tools is equally taken up by proponents of a human rights frameworks based approach with the expectation that these can be adapted to ensure an integration of human rights norms within the AI lifecycle (Yeung et al. 2019). Accompanying methods and measures stand to strengthen Actionable Principles by providing guidance as to how their recommendations should be implemented and by whom. Moreover, recent academic work on mechanisms for supporting verifiable claims (Brundage et al. 2020) highlighted the importance of providing such mechanisms to actors involved throughout the AI lifecycle, including governments.

In order to account for the broad impact of AI systems it is necessary to consider and provide both technical and non-technical tools and measures. Indeed, the Ethics Guidelines did both. Technical measures could for example cover explainable AI research (XAI), which can help to illuminate some of the underlying decision making processes within some AI systems, increasing an individual's ability to understand (and challenge) the AI system's output. They could also entail privacy-preserving measures, which serve to adequately protect and secure (personal) data used to develop and maintain a given AI system's functionality; requirements for sufficiently representative data sets (Flournoy et al. 2020) to ensure both adequate functioning of an AI system in a real-world environment as well as to tackle bias introduced through inadequate data sets; or, describe testing and validation protocols and procedures that would support the fulfillment of the principles in question by a given AI system. The latter may cover adversarial testing by dedicated red teams (Brundage et al. 2020), evaluation of out of distribution robustness (Lohn 2020), repeat testing within reasonable time spans once deployed, or denoting adequate requirements for audit trails at all stages of the lifecycle of a given AI system (AI

HLEG 2020).

Non-technical (or less technical) measures can provide requirements and methods to increase civil society's participation in decision-making processes surrounding the development and deployment of AI systems, empowering them to (safely) call out and halt the development and application of AI systems with potentially negative impacts. The dialogue with civil society and affected parties could be strengthened through e.g. the development and provision of relevant 'algorithmic impact assessments' (Reisman et al. 2018). An example of this could be the Canadian government's algorithmic impact assessment[9] or the AI HLEG's revised assessment list for trustworthy AI (AI HLEG 2020). Other measures could cover hiring practices, institutional mechanisms whereby employees can safely flag concerns surrounding a given AI system (e.g. its performance or scope), or the creation of audit trails (e.g. to enable the auditing of AI systems by third parties) (Raji et al. 2020). Moreover, other relevant organisational efforts such as incentivising AI developers through codes of ethics or codes of conduct could be suggested.

Finally, as stated earlier, Actionable Principles may influence policy-making simply by virtue of shaping governmental funding decisions. Along these lines, principles that concern e.g. the protection of the environment and a support for the flourishing of future generations (AI HLEG 2019b) could benefit from being accompanied by recommendations for funding research into methods such as computationally efficient algorithms (Strubell et al. 2019).

While the provision of mechanisms for implementation does not replace a clear and coherent structure of the document in question, it can contribute to an actionability of underlying goals. Furthermore, this directly builds and expands on the other two elements: the landscape assessment and multi-stakeholder engagement strategy. Together forming a prototype framework across the lifecycle of Actionable Principles—inception, development, post-publication.

---

[9] https://open.canada.ca/aia-eia-js/?lang=en.

As discussed earlier, a prominent critique of AI Ethics Principles at large is that they are susceptible to 'ethics washing', including concerns that they may run risk to morph away from moral principles into a 'performative facade' (Bietti 2019) and, ultimately, that they are an 'easy' or 'soft' option (Wagner 2018) in comparison to 'hard(er)' governance mechanisms such as regulation. This paints them as a distraction rather than a solution to the problems they are hoping to address. In fact, even members of AI HLEG have criticized the Ethics Guidelines over 'ethics washing', alleging excessive industry influence (Metzinger 2019). It should be noted that these criticisms are largely launched against the role of industry in operationalizing AI Ethics (Ochigame 2019) or developing AI Ethics Principles in order to forestall or evade government regulation and real oversight. The focus of this paper, however, is on governments and their policy-making, i.e. on *increasing* the influence Actionable Principles can have on concrete soft and hard governance developments. It sees ethical enquiry as key to develop good policy even if the focus of the framework is currently limited to procedural aspects as enablers.

In conclusion, it is suggested that the provision of methods and mechanisms to increase the actionability of recommendations in Actionable Principles through e.g. testing and validation protocols, cross-societal dialogue and audit trails further closes the gap by providing tools to move from principles to policy.

## Conclusion

While their rise has been encouraging and needed, the majority of AI Ethics Principles today still suffer from a lack of actionability in policy-making. This paper has suggested that instead of abandoning the approach altogether, the community should iterate on and improve these tools in order to produce a form of 'Actionable Principles' on AI, which integrate actionability and ethical reasoning. In order to do so, this paper first briefly identified and examined a series of promising elements from the development process of the Ethics Guidelines by the AI HLEG. Subsequently, it proposed three elements towards a framework for Actionable Principles: (1)

preliminary landscape assessments; (2) multi-stakeholder participation and cross-sectoral feedback; and, (3) mechanisms to support implementation and operationalizability.

Given the current pace of governmental initiatives and the drastically increasing number of groups working to develop relevant guidance, this paper's suggestion comes at a crucial time. Excitement about AI Ethics Principles has waned as the reality has set in that significant work from all stakeholders needs to be undertaken to move them from paper to practice. Multiple efforts are underway to do so (e.g. via the Global Partnership on AI's committee on responsible AI[10] or the German government's project on 'ethics of digitalisation'[11]) and it is hoped that the prototype framework provided in this paper sets out a useful path. Inevitably, actionability is not and will not be the only hurdle AI Ethics Principles will face over the coming years. Nevertheless, the pacing-problem (Marchant et al. 2011) is real and future governance efforts will significantly rely on existing (academic) work as they make sense of direly needed policy options. This puts Actionable Principles in a potentially powerful political and societal role, one that needs to be taken seriously and nourished.

**Acknowledgements:** The author would like to thank Vincent C. Müller, Michael Cannon, Gabriela Arriagada Bruneau, Matthijs M. Maas, Hiski Haukkala and three anonymous reviewers for commentary and suggestions on previous versions of this paper.

---

[10] https://oecd.ai/wonk/an-introduction-to-the-global-partnership-on-ais-work-on-responsible-ai

[11] https://www.bundespraesident.de/SharedDocs/Berichte/DE/Frank-Walter-Steinmeier/2020/08/200817-Ethik-der-Digitalisierung.html